\documentclass{elsart}

\usepackage{graphicx}

\newcommand{\vspfigA}{\vspace{0cm}}  
\newcommand{\vspfigB}{\vspace{0cm}} 
\newcommand{\vspfigC}{\vspace{1cm}}
\newcommand{\widthfigA}{0.95\textwidth} 
\newcommand{\widthfigB}{0.9\textwidth} 
\newcommand{\widthfigC}{0.5\textwidth} 
\newcommand{\sectionspace}{\vspace{0.5cm}}

\begin{document} 

%%%%%%%%%%%%%%%%%%%%%%%%%%%%%%%%%%%%%%%%%%%%%%%%%%%%%%%%%%%%%%%%%%%%%%

\begin{frontmatter}
 
\title{Lyapunov Modes and Time-Correlation Functions
for Two-Dimensional Systems}  
 
\author{Tooru Taniguchi and Gary P. Morriss}

\address
%\affiliation
{School of Physics, University of New South Wales, 
   Sydney, New South Wales 2052, Australia} 
\date{\today}

%\maketitle
%\vspace{0cm} 
\begin{abstract}

%\linesabs
      The relation between the Lyapunov modes (delocalized 
   Lyapunov vectors) and the momentum autocorrelation function 
   is discussed in two-dimensional hard-disk systems. 
      We show numerical evidence that 
   the smallest time-oscillating period of the Lyapunov modes 
   is twice as long as the time-oscillating period of 
   momentum autocorrelation function for 
   both square and rectangular two-dimensional systems 
   with hard-wall boundary conditions. 

%   \vspace{0.5cm}\noindent \mbox{\bf Keywords:}

\end{abstract} 


\end{frontmatter} 
%\maketitle

%%%%%%%%%%%%%%%%%%%%%%%%%%%%%%%%%%%%%%%%%%%%%%%%%%%%%%%%%%%%%%%%%%%%%%

\section{Introduction}

   The Lyapunov mode is a delocalized structure appearing 
in the Lyapunov vectors corresponding to the Lyapunov exponents 
close to zero in many-body chaotic systems. 
   The existence of such a mode structure for Lyapunov vectors 
was first suggested from the stepwise structure of the Lyapunov 
spectra for many-hard-core-particle systems \cite{Del96,Mil98a}, 
and led to the discovery of a stationary 
%translational 
transverse 
T-mode structure in the Lyapunov vectors \cite{Mil98a,Pos00,Mil02}. 
   Further investigations clarified the other 
two kinds of Lyapunov modes, 
the longitudinal L-mode  \cite{For04} 
and the momentum proportional 
P-mode \cite{Tan03a,Tan05a,Tan05b}, both of 
which contain an explicit time-dependence. 
   Combining these three kinds of Lyapunov modes,  
explained the degeneracies in the stepwise structure of 
Lyapunov spectra \cite{Tan05a,Tan05b,Eck05}. 
   Moreover, the 
structure of these Lyapunov modes 
suggests that the origin of the Lyapunov modes is 
the dynamical conservation laws,  
like energy and momentum conservation,  
and spatial and temporal translational invariances 
\cite{Tan03a,Tan05a,Tan05b,Eck05} 
and Noether's theorem.
   The conservation laws and translational invariances 
dominate the global and thermodynamic behavior of many-body systems, 
and the behavior of Lyapunov vectors for the 
Lyapunov exponents close to zero should correspond to 
slow and collective movements. 
  Therefore it is expected that the Lyapunov modes (which are chaotic properties)
allow us to discuss the 
thermodynamic behavior of many-body systems dynamically. 
   Many studies of the Lyapunov modes have used  
hard-core-particles in quasi-one-, two- or three-dimensional systems 
because of its established 
Lyapunov vector dynamics and clear mode structures.  
   Recently Lyapunov modes 
for soft-core particle systems  
%and coupled map systems, 
have also been investigated \cite{Hoo02,For04b,Yan04}. 
   Some analytical approaches have been proposed 
to understand the Lyapunov modes and the
stepwise structure of Lyapunov spectra. 
   These include  
a random matrix approach \cite{Eck00}, 
a Fokker-Planck equation approach \cite{Tan02c}, 
kinetic approaches \cite{Mcn01b,Mar04,Wij04}
and a periodic orbit approach \cite{Tan02b}.

   One recent development in the study  
of Lyapunov modes is the discovery of a connection between 
the time dependence of the Lyapunov modes and the momentum 
autocorrelation function \cite{Tan05a,Tan05b}.
   The integral of the momentum autocorrelation function gives 
the transport coefficient (here the diffusion coefficient) 
through linear response theory \cite{Kub78}. 
   Moreover it includes information on 
the collective movement of many-body systems 
\cite{Zwa67,Rah67,Han90}, 
and is accessible experimentally using neutron and light 
scattering techniques \cite{Cop75,Wel85,Sch99}.
   It was shown in Ref. \cite{Tan05a,Tan05b} 
that the largest time-oscillating period $T_{Lya}$ of 
time-dependent Lyapunov modes is twice as long as 
the time-oscillating period $T_{tcf}$ 
of the momentum autocorrelation function, namely 
\begin{eqnarray}
   T_{Lya} = 2 T_{tcf}. 
\label{PerioRelat}\end{eqnarray}
   This relation connects the Lyapunov mode and an 
experimentally accessible quantity.  
   In reference \cite{Tan05a,Tan05b}, 
a theoretical argument to support this relation, and 
numerical evidence that this relation 
is independent of density, particle number, and boundary 
conditions in quasi-one-dimensional systems was given. 
%   However, they did not show numerical evidence 
%that this relation 
%is independent of the spatial dimension of the system.   
%
   The aim of this paper is to present numerical evidence 
that the relation (\ref{PerioRelat}) is satisfied 
not only in quasi-one-dimensional systems but also 
in fully two-dimensional systems.

   To calculate Lyapunov exponents and Lyapunov vectors 
we use the Lyapunov vector dynamics for hard-disk systems 
\cite{Del96}, 
and the algorithm developed by Benettin \textit{et al.} 
\cite{Ben80a}, and Shimada and Nagashima \cite{Shi79}. 
   In general, the difference between two nearby 
phase space trajectories defines a direction in phase space.
   The rate of growth of this difference is controlled by the 
largest Lyapunov exponent, and the normalized direction (which is
time dependent) is the Lyapunov vector for that exponent. 
   Eliminating the direction defined by the Lyapunov vector for the 
largest exponent, reduces the dimension of the phase space by
one.
   In that reduced phase space we can again define a direction
that corresponds to the (next) largest Lyapunov exponent
and its associated Lyapunov vector.
   Continuing this procedure we define the same number of  
Lyapunov exponents 
and Lyapunov vectors as there are independent directions
in phase space. 
   This process is essentially the Benettin scheme 
\cite{Ben80a} for calculating the full spectrum
of Lyapunov exponents for the system.
   The only additional element is the significance given to basis vectors
used in the Benettin scheme; here they called Lyapunov vectors
and become an object of study in their own right.
   While the Lyapunov vectors associated with the largest exponents are
strongly localized (having non-zero components for only a few particles)
the vectors associated with the smallest positive and negative exponents
have delocalized Lyapunov vectors that we refer to as {\it Lyapunov modes}.

%\pagebreak   
     
%%%%%%%%%%%%%%%%%%%%%%%%%%%%%%%%%%%%%%%%%%%%%%%%%%%%%%%%%%%%%%%%%%%%%%
%\sectionspace
\section{A Square System with 100 Disks}

%---------------------------------------------------------------------
\begin{figure}[!t]
\vspfigA
\begin{center}
\includegraphics[width=\widthfigC]{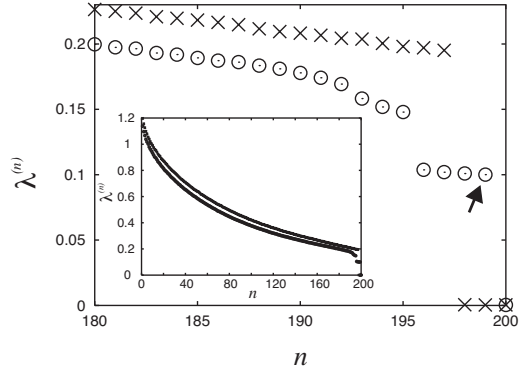}
\end{center}
\vspfigB
\caption{
      The stepwise structure of Lyapunov spectrum for a square system 
   consisting of $100$ hard-disks with hard-wall boundary conditions 
   (circles). 
      For a comparison, the Lyapunov spectrum for a square system 
   with periodic boundary conditions (crosses) is also shown.
      The Lyapunov exponent indicated by an arrow is 
   the one in which a time-oscillation of Lyapunov mode 
   is shown in Fig. \ref{fig2modeOscil}(b). 
      Inset: Full positive branch of Lyapunov spectra. 
%
%      (b) The time-oscillation of the locally spatial-temporal averaged 
%   Lyapunov vector component 
%   $\langle\delta y^{(2N-1)}\rangle_{t}$ 
%   for 
%   %the $y$-component 
%   the component $dy_{j}^{(2N-1)}$ 
%   %of the spatial part 
%   of the Lyapunov vector. 
%      The line is a fit of numerical data to a sinusoidal 
%   function. 
   }
\vspfigC
\label{fig1lyapuSqu}\end{figure}  
%-------------------------------------------------------------------- 
%
   We investigate 
the time-oscillating periods of the Lyapunov modes and 
the momentum autocorrelation function, 
for a system of hard-disks in a square 
with the hard-wall boundary conditions. 
   Figure \ref{fig1lyapuSqu} is the Lyapunov spectrum for 
this system (circles). 
   Here, we used as system parameters: 
$N=100$, disk radius $R=1$, mass $M=1$, 
total energy $E=N$, and the system lengths in the $x$ and
$y$ directions $L_{x}=L_{y}\approx 38.7$, 
so that the particle number density is 
$N(2R)^2/(L_{x}L_{y})\approx 0.267$. 
   The Lyapunov spectrum shows a 4-point step close to the 
two zero-Lyapunov exponents. 
   We use hard-wall boundary conditions because  
that leads to more and clearer steps in the Lyapunov spectrum 
than for the equivalent system with periodic boundary conditions. 
   In fact we do not observe stepwise structure of the 
Lyapunov spectrum for a 100-disk system  
with periodic boundary conditions.
   To show this, in Fig. \ref{fig1lyapuSqu}, 
we also plotted the Lyapunov spectrum for 
a square system with periodic boundary conditions (crosses).
   (Here we used the same system parameters 
as for hard-wall boundary conditions, except that  
$L_{x}=L_{y}\approx 36.7$ so the effective region 
available for the disks to move is the same in both cases.) 
%   Different from the case of the hard-wall boundary conditions, 
   The system with periodic boundary conditions is too small 
to observe the stepwise structure of the Lyapunov spectrum. 
   The number of zero-Lyapunov exponents is 2 for hard-wall 
boundary conditions, because of energy conservation 
and time-translational invariance (determinisity of orbit), 
and is 6 for periodic boundary conditions, 
because of total momentum conservation, 
conservation of center of mass in both directions, 
energy conservation and time translational invariance 
\cite{Gas98}. 
   Only half of those zero-Lyapunov exponents are presented in  
Fig.  \ref{fig1lyapuSqu}. 
  
%---------------------------------------------------------------------
\begin{figure}[!t]
\vspfigA
\begin{center}
\includegraphics[width=\widthfigA]{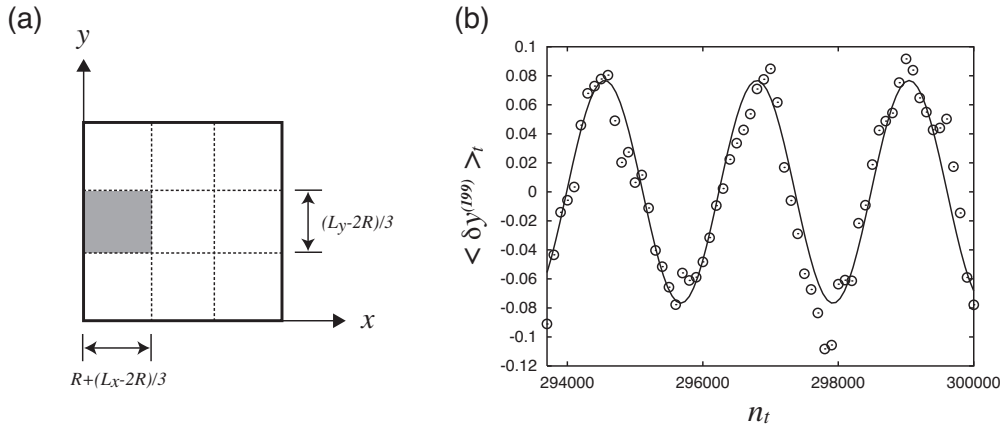}
\end{center}
\vspfigB
\caption{
%      Time-oscillation of Lyapunov mode. 
      (a) The spatial region of the system in which 
   the average of the longitudinal Lyapunov mode of 
   Fig. (b), is taken.  
%   the $x$ and $y$-axes shown. 
      (b) The time-oscillation of the local spatial region for the  
   Lyapunov vector component 
   $\langle\delta y^{(2N-1)}\rangle_{t}$ 
   for the component $dy_{j}^{(2N-1)}$ 
   of the Lyapunov vector. 
      The line is a fit of the numerical data to a sinusoidal 
   function. 
      The system is the same as that in Fig. \ref{fig1lyapuSqu}
    with hard-wall boundary conditions.  
%      This Lyapunov vector correspond to the Lyapunov 
%   exponent $\lambda^{(2N-1)}$ indicated by the arrow 
%   in Fig. \ref{fig1lyapuSqu}.  
   }
\vspfigC
\label{fig2modeOscil}\end{figure}  
%-------------------------------------------------------------------- 
%  
   We are interested in the time-oscillation of modes 
corresponding to the stepwise 
structure of the Lyapunov spectrum shown in Fig. \ref{fig1lyapuSqu}. 
   It is known that there are two kinds of 
time-dependent Lyapunov modes, the L-modes and 
the P-modes, each with the same period  
\cite{Tan03a,Tan05a,Tan05b,Eck05}.
   Our concern is to determine the time-oscillating period 
of Lyapunov mode, so it is enough to consider one of these.
   We choose the L-mode 
and concentrate into its time-oscillating behavior only, 
ignoring the spatial structure \cite{Tan05a,Tan05b,Eck05}. 
   To determine the period of the L-mode 
we use the same system with hard-wall boundary conditions,  
see Fig. \ref{fig1lyapuSqu}, and consider 
the $\delta y_{j}^{(2N-1)}$. 
   This Lyapunov vector component $\delta y_{j}^{(2N-1)}$ 
corresponds to the Lyapunov exponent $\lambda^{(2N-1)}$  
and is indicated by the arrow in Fig. \ref{fig1lyapuSqu}. 
   To get clear time-oscillating behavior 
we take the average of $\delta y_{j}^{(2N-1)}$ 
for disks whose centers $(x_{j},y_{j})$ are in the region 
$R < x_{j}<R+(L_{x}-2R)/3$ 
and $R+(L_{y}-2R)/3 < y_{j} < R+2 (L_{y}-2R)/3$  
(the grey region in Fig. \ref{fig2modeOscil}(a)) 
at the time of the collisions. 
   This region is chosen because the L-mode 
has a node at this end of the system, so that 
the longest L-mode for $\delta y_{j}^{(2N-1)}$ 
should have nodes at $y=R$ and $L_{y}-2R$ and 
an anti-node at $y=L_{y}/2$, so 
the Lyapunov mode $\delta y_{j}^{(2N-1)}$ should have  
largest amplitude.
   Further, we take a local-time average of the  
spatially averaged Lyapunov vector component $\delta y_{j}^{(2N-1)}$ 
over $N$ successive disk-disk collisions.  
   This kind of local spatial-temporal average 
is required to clearly observe the time-oscillation of Lyapunov modes.  
  A square system of $N=100$ particles 
is not large enough to observe clear mode structures in the  
Lyapunov vectors, as there are only about $\sqrt{N}=10$ 
particles in each coordinate direction. 
   Fig. \ref{fig2modeOscil} is a graph of the  
local spatial-temporal average $\langle\delta y^{(2N-1)}\rangle_{t}$ 
of $dy_{j}^{(2N-1)}$ as a function of the collision number $n_{t}$. 
   In this figure we can observe the time-oscillation of 
the longitudinal Lyapunov mode, and the fit of 
the graph to a sinusoidal function 
$\langle\delta y^{(2N-1)}\rangle_{t} 
= \alpha \sin(2\pi n_{t}/T_{Lya} +\beta)$ 
with the fitting parameters $\alpha$, $\beta$ and $T_{Lya}$.  
   This gives its time-oscillating period 
$T_{Lya}$ as $T_{Lya}\approx 2250$. 
%A1 = 0.0767524
%B1 = 2247.48
%C1 = 1.20364
   Note that $T_{Lya}$ is in units of collision numbers 
$n_{t}$, and the real time interval of the time-oscillating period of 
this Lyapunov mode is given by $\tau T_{Lya}$ 
where $\tau \approx 0.0279$ is the mean free time. 
%0.0279815 

%---------------------------------------------------------------------
\begin{figure}[!t]
\vspfigA
\begin{center}
\includegraphics[width=\widthfigB]{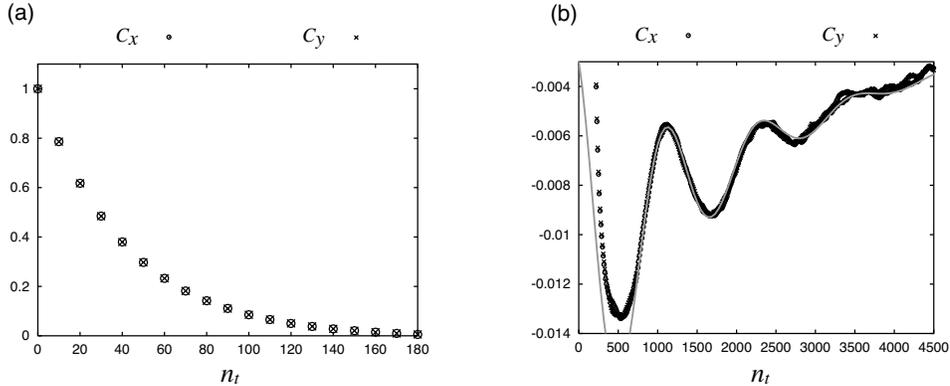}
\end{center}
\vspfigB
\caption{
      The momentum autocorrelation functions 
%   [its $x$-component is $C_{x}$ (circles) 
%   and its $y$-component is $C_{y}$ (crosses)] 
   as a function of the collision number $n_{t}$ 
   for a square system consisting of 100 disks. 
      (a) The initial damping behavior of 
   the momentum autocorrelation functions. 
   %$C_{x}$ and $C_{y}$ 
   %  as a function of the collision number $n_{t}$. 
      (b) The time-oscillating region of 
   the momentum autocorrelation functions. 
      %$C_{x}$ and $C_{y}$.  
      $C_{x}$ (circles) is the autocorrelation function of the $x$-component 
   and $C_{y}$ (crosses) is the autocorrelation function of the $y$-component.  
      These two components should be equal for a square system. 
      The system is the same as that in Fig. \ref{fig2modeOscil}.  
      The line in Fig. (b) is a fit of  $C_{x}$ to an exponentially damped 
      sinusoidal function.
   }
\vspfigC
\label{fig3tcfOscil}\end{figure}  
%-------------------------------------------------------------------- 
%
   Next, we consider the time-oscillating behavior 
of the momentum autocorrelation function. 
   For this purpose we introduce 
the autocorrelation functions $C_{\eta} (t)$ of the components of momentum 
(where $\eta =x$ or $y$) 
and use the normalized expression 
$C_{\eta}(t) \equiv \tilde{C}_{\eta}(t)/\tilde{C}_{\eta}(0)$, 
in which $\tilde{C}_{\eta}(t)$ is defined by 
%
%\begin{eqnarray}
$
   \tilde{C}_{\eta}(t) \equiv 
      \lim_{T\rightarrow +\infty} 
      \frac{1}{NT} 
   %\nonumber \\
   %&& \times 
   \sum_{j=1}^{N} 
   \int_{0}^{T} ds \; 
      p_{\eta j}(s+t) p_{\eta j}(s).   
$
%\label{CorreFunct}\end{eqnarray}
%  
Here $p_{\eta j}(s)$ is the $\eta$-component of the 
momentum of the $j$-th disk.  
   (In numerical calculations, the time-average 
for the time $s$ in the above definition of autocorrelation 
function is replaced 
by the arithmetic average of $p_{\eta j}(s+t) p_{\eta j}(s)$ 
after each collision.)  
   Figure \ref{fig3tcfOscil} shows the autocorrelation 
functions $C_{x}$ and $C_{y}$ 
for the $x$- and $y$-components of momenta, respectively,  
as functions of the collision number $n_{t}$ in the same system 
as that used in Fig. \ref{fig2modeOscil}. 
   The autocorrelation functions $C_{x}$ and $C_{y}$ 
coincide as they should for a square system. 
   Fig. \ref{fig3tcfOscil}(a) shows the initial  
damping behavior of the autocorrelation functions, 
and Fig. \ref{fig3tcfOscil}(b) is 
the negative region in which the 
time-oscillating behavior of the autocorrelation functions appears. 
   We can extract the time-oscillating period $T_{tcf}$ 
by fitting this region of the  
autocorrelation function to the function 
$C_{x} = \alpha \exp\{-\beta n_{t}\} + \gamma 
\exp\{-\epsilon n_{t}\} \sin (2\pi n_{t}/T_{tcf} + \zeta)$ 
(the line in Fig. \ref{fig3tcfOscil}(b)) 
with the fitting parameters $\alpha$, $\beta$, $\gamma$, 
$\epsilon$ and $\zeta$. 
  This fit gives the value of the period for the momentum 
autocorrelation function as $T_{tcf}\approx 1190$. 
   Comparison of the value $T_{tcf}$ 
with the longest time-oscillating period $T_{Lya}$ 
for the Lyapunov modes gives further evidence to support  
the relation (\ref{PerioRelat}) in two-dimensional systems.

%%%%%%%%%%%%%%%%%%%%%%%%%%%%%%%%%%%%%%%%%%%%%%%%%%%%%%%%%%%%%%%%%%%%%%
%\sectionspace
\section{A Rectangular System with 780 Disks}
  
%---------------------------------------------------------------------
\begin{figure}[!t]
\vspfigA
\begin{center}
\includegraphics[width=\widthfigA]{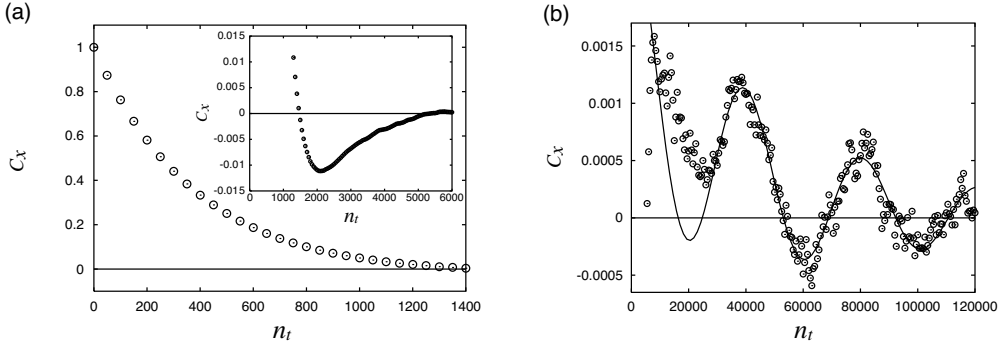}
\end{center}
\vspfigB
\caption{
      The momentum autocorrelation function $C_{x}$ 
   as a function of the collision number $n_{t}$ 
   in a rectangular system consisting of 780 disks.  
      (a) The initial damping of $C_{x}$ for $0<n_{t}<1400$.    
      Inset: The negative region of $C_{x}$ for $1500<n_{t}<6000$.
      (b) The time-oscillating part of $C_{x}$ for $10000<n_{t}<120000$. 
      The line is a fit of 
%   the time-oscillating part of 
   $C_{x}$ to an exponentially decaying sinusoidal function.
   }
\vspfigC
\label{fig4tcfRec780}\end{figure}  
%-------------------------------------------------------------------- 
%
   A system of 100 particles is not sufficient to investigate 
the Lyapunov modes in two-dimensions as 
only a few steps of Lyapunov spectrum are observed,
even for hard-wall boundary conditions.  
   A larger two-dimensional system of 780 particles  
with hard-wall boundary conditions 
has already been investigated in Ref. \cite{Eck05}
and the Lyapunov modes calculated. 
   The system parameters used were: particle mass $M=1$, 
disk radius $R=1/2$, aspect ratio $L_{y}/L_{x} = 0.867$, 
and particle number density $N(2R)^{2}/(L_{x}L_{y}) = 0.8$. 
   In this system the observed time-oscillating period 
for the Lyapunov mode is $T_{Lya}\tau \approx 12.80$, 
namely $T_{Lya} \approx 74000$ collisions numbers 
with the mean free time of $\tau \approx 0.000173$. 
%73988.439306358

   Figure \ref{fig4tcfRec780} is the normalized 
autocorrelation function $C_{x}$
of the $x$-component of the momentum as a function 
of the collision number $n_{t}$. 
   Here, the system shape is rectangular and  
$L_{x}>L_{y}$. 
   This means that the time-oscillating period of 
$C_{x}$ is larger than that for $C_{y}$, 
and should correspond to the $T_{Lya}$ observed in Ref. \cite{Eck05}. 
   The autocorrelation function $C_{x}$, normalized by 
its initial value, shows an initial exponential damping, 
as shown in Fig. \ref{fig4tcfRec780}(a), and after that 
a negative region appears, as in Fig. \ref{fig3tcfOscil} 
for the $N=100$ case. 
   The time-oscillating behavior of $C_{x}$ appears after 
the negative region of $C_{x}$, as shown in Fig.  
\ref{fig4tcfRec780}(b). 
   To extract the time-oscillating period $T_{tcf}$ for $C_{x}$ 
we fitted this time-oscillating region of the 
autocorrelation function to the function 
$C_{x} = \alpha \exp\{-\beta n_{t}\} + \gamma 
\exp\{-\epsilon n_{t}\} \sin (2\pi n_{t}/T_{tcf} + \zeta)$  
with the fitting parameters $\alpha$, $\beta$, $\gamma$, 
$\epsilon$ and $\zeta$.
   This fit gives us a value of $T_{tcf}\approx 40800$ 
for time-oscillating period of $C_{x}$. 
   This result again confirms that  
the relation (\ref{PerioRelat}) is satisfied approximately
for two-dimensional systems. 
%A1= 0.00126571
%B1= 40754.3
%C1= 6.37054
%D1= 1.35623e-05
%A2= 0.00151925
%B2= 3.40528e-05

   We emphasize that this time-oscillating 
behavior of the autocorrelation function $C_{x}$ 
appears much later than its initial damping, 
and with a very small amplitude. 
   (The above fit gives 
$\gamma\approx 0.00127$ as the amplitude 
of the time-oscillation of $C_{x}$ shown in Fig. 
\ref{fig4tcfRec780}(b), 
which is much smaller than the initial value $1$ for $C_{x}$.)
   Besides, the time-oscillation of 
the autocorrelation function has the time-scale much longer than 
its decay time.
   This means that very long numerical calculations are required  
to obtain the time-oscillating behavior of $C_{x}$ in such a large system, 
and we calculated more than $5\times 10^{7}$ 
collisions to get the data presented in Figs. 
\ref{fig3tcfOscil}(b) and \ref{fig4tcfRec780}(b). 
  These are much longer than the simulations required 
to calculate Lyapunov spectra.   
  
%%%%%%%%%%%%%%%%%%%%%%%%%%%%%%%%%%%%%%%%%%%%%%%%%%%%%%%%%%%%%%%%%%%%%%
%\sectionspace
%\vspace{3cm}
\section{Conclusion}

   In conclusion, this paper gives further numerical evidence 
that the relation $T_{Lya} \approx 2 T_{tcf}$ 
for the smallest time-oscillating period $T_{Lya}$ 
of the Lyapunov mode and the time-oscillating period 
$T_{tcf}$ of the momentum autocorrelation function 
is satisfied in fully two-dimensional hard-disk systems. 
   This supports our previous results reported in 
Ref. \cite{Tan05a,Tan05b}, adding numerical evidence 
that the relation is independent of the spatial dimension 
of the system. 

   As a remark, in this paper we used a simple fit 
of time-oscillation of the autocorrelation function and Lyapunov modes 
to functions including a sinusoidal function of time 
in order to extract their time-oscillating periods, 
as was done for the quasi-one-dimensional systems \cite{Tan05a,Tan05b}. 
   However, the amplitude of the time-oscillating of 
momentum autocorrelation function is smaller for two-dimensional systems 
than for quasi-one-dimensional systems. 
  Considering the Fourier transformation of momentum autocorrelation 
function and Lyapunov modes may be more efficient technique
to extracting the periods.

%%%%%%%%%%%%%%%%%%%%%%%%%%%%%%%%%%%%%%%%%%%%%%%%%%%%%%%%%%%%%%%%%%%%%%
\sectionspace

   The authors of this paper appreciate the financial support 
of the Japan Society for the Promotion of Science.

%%%%%%%%%%%%%%%%%%%%%%%%%%%%%%%%%%%%%%%%%%%%%%%%%%%%%%%%%%%%%%%%%%%%%%
\sectionspace

%%%%%%%%%%%%%%%%%%%%%%%%%%%%%%%%%%%%%%%%%%%%%%%%%%%%%%%%%%%%%%%%%%%%%%


\begin{thebibliography}{00} 
\vspace{-0.5cm}

%--- Lyapunov mode 

\bibitem{Del96} Ch. Dellago, H. A. Posch, and W. G. Hoover, 
   Phys. Rev. E {\bf 53}, 1485 (1996).
\bibitem{Mil98a} Lj. Milanovi\'c, H. A. Posch, and Wm. G. Hoover, 
Mol. Phys. {\bf 95}, 281 (1998).
\bibitem{Pos00} H. A. Posch and R. Hirschl, 
   in {\it Hard ball systems and the Lorentz gas}, 
   edited by D. Sz\'asz (Springer-Verlag, Berlin, 2000), 
   p. 279. %EMS Vol 101, 
\bibitem{Mil02} Lj. Milanovi\'c and H. A. Posch, 
   J. Molec. Liquids, {\bf 96-97}, 221 (2002). 
\bibitem{For04} C. Forster, R. Hirschl, H. A. Posch, 
   and W. G. Hoover, Physica D  \textbf{187}, 294 (2004). 
\bibitem{Tan03a} T. Taniguchi and G. P. Morriss, 
   Phys. Rev. E \textbf{68}, 026218 (2003).
\bibitem{Tan05a} T. Taniguchi and G. P. Morriss
   Phys. Rev. E \textbf{71} 016218 (2005). 
\bibitem{Tan05b} 
   T. Taniguchi and G. P. Morriss, 
   Phys. Rev. Lett. \textbf{94}, 154101 (2005).
\bibitem{Eck05} J. -P. Eckmann, C. Forster, H. A. Posch,  
   and E. Zabey, 
   J. Stat. Phys. \textbf{118}, 813 (2005).
   
\bibitem{Hoo02} Wm. G. Hoover, H. A. Posch, C. Forster, 
   C. Dellago, and M. Zhou, 
   J. Stat. Phys., \textbf{109}, 765 (2002).  
\bibitem{For04b} 
   C. Forster and H. A. Posch, 
   New J. Phys. \textbf{7}, 32 (2005).
\bibitem{Yan04} 
   H. Yang and G. Radons, 
   Phys. Rev. E \textbf{71}, 036211 (2005).
%\bibitem{Rad04} 
%   G. Radons, 
%   in the workshop "Stochastic and Deterministic Dynamics 
%   in Equilibrium and Nonequilibrium Systems," 
%   The International Erwin Schringer Institute (ESI), 
%   Vienna, Austria, August, 2004. 


%--- Lyapunov step and mode (analytical)

\bibitem{Eck00} J. -P. Eckmann and O. Gat, 
   J. Stat. Phys. \textbf{98}, 775 (2000).
\bibitem{Tan02c} T. Taniguchi and G. P. Morriss, 
   Phys. Rev. E \textbf{65}, 056202 (2002).
\bibitem{Mcn01b} S. McNamara and M. Mareschal, 
   Phys. Rev. E \textbf{64}, 051103 (2001).
\bibitem{Mar04} 
   M. Mareschal  and S. McNamara,  
   Physica D \textbf{187}, 311 (2004).  
\bibitem{Wij04} 
   A. S. de Wijn and H. van Beijeren, 
   Phys. Rev. E \textbf{70}, 016207 (2004).
\bibitem{Tan02b} T. Taniguchi, C. P. Dettmann, and G. P. Morriss, 
   J. Stat. Phys. \textbf{109}, 747 (2002). 


%--- Linear response theory 

\bibitem{Kub78} R. Kubo, M. Toda, and N. Hashitsume, 
   \textit{Statistical physics II, 
   nonequilibrium statistical mechanics} 
   (Springer-Verlag, Berlin, 1985).   

   
%--- Time-oscillation of the auto-correlation function

\bibitem{Zwa67} R. Zwanzig, 
   Phys. Rev. \textbf{156}, 190 (1967).
%  
\bibitem{Rah67} A. Rahman, 
   Phys. Rev. Lett. \textbf{19}, 420 (1967);
W. E. Alley, B. J. Alder, and S. Yip, 
   Phys. Rev. A \textbf{27}, 3174 (1983).
\bibitem{Han90} 
   J. P. Hansen and I. R. McDonald, 
   \textit{Theory of simple liquids}, 2nd ed. 
   (Academic press, London, 1990). 
%\bibitem{Ber81} M. Berkowitz and J. A. McCammon, 
%   J. Chem. Phys. \textbf{75}, 957 (1981).


%--- Light scattering technique for the autocorrelation function

\bibitem{Cop75} J. R. D. Copley and S. W. Lovesey, 
   Rep. Prog. Phys. \textbf{38}, 461 (1975). 
\bibitem{Wel85} 
   A. A. van Well, P. Verkerk, L. A. de Graaf,  
   J.-B. Suck, and J. R. D. Copley, 
   Phys. Rev. A \textbf{31}, 3391 (1985). 
%\bibitem{Sch98} W. Sch\"artl, C. Roos, and K. Gohr, 
%   J. Chem. Phys. \textbf{108}, 9594 (1998).
\bibitem{Sch99} 
   W. Schaertl and C. Roos,
   Phys. Rev. E \textbf{60}, 2020 (1999). 
   


%--- Numerical Method for computation of Lyapunov spectra

%\bibitem{Ben76} G. Benettin, L. Galgani, and J. -M. Strelcyn, 
%   Phys. Rev. A \textbf{14}, 2338 (1976).
%\bibitem{Ben78} G. Benettin, L. Galgani, A. Giorgilli 
%   and J. M. Strelcyn, C. R. Acad. Sci. Ser. A {\bf 286}, 
%   431 (1978).
\bibitem{Ben80a} G. Benettin, L. Galgani, A. Giorgilli, 
   and J. -M. Strelcyn, 
   Meccanica \textbf{15}, 9 (1980); 
%\bibitem{Ben80b} G. Benettin, L. Galgani, A. Giorgilli, 
%   and J. -M. Strelcyn, 
   Meccanica \textbf{15}, 21 (1980).
\bibitem{Shi79} I. Shimada and T. Nagashima, 
   Prog. Theor. Phys. \textbf{61}, 1605 (1979).

%---

\bibitem{Gas98} P. Gaspard, Chaos, Scattering and Statistical 
   Mechanics (Cambridge University Press, Cambridge, England, 1998). 
  
\end{thebibliography}
\end{document}